\documentclass[12pt]{iopart}

\usepackage{graphicx}
\usepackage{color}
\usepackage{amssymb}

\usepackage[implicit=true,colorlinks=true,linkcolor=blue,
citecolor=blue,urlcolor=blue]{hyperref} 

\newcommand{\dN}{64}
\newcommand{\dFa}{15}
\newcommand{\dFb}{35}
\newcommand{\La}{10}
\newcommand{\Lb}{17}

\newcommand{\dd}{\mathrm{d}}
\newcommand{\ii}{\mathrm{i}}
\newcommand{\ee}{\mathrm{e}}
\newcommand{\sign}{\mathrm{sign}}
\newcommand{\omegat}{\tilde{\omega}}
\newcommand{\Omegat}{\widetilde{\Omega}}

\newcommand{\Tc}{T_\mathrm{c}}
\newcommand{\dFt}{d_\mathrm{F}}
\newcommand{\dNt}{d_\mathrm{N}}

\newcommand{\gBSN}{\gamma_\mathrm{BSN}}
\newcommand{\gBSF}{\gamma_\mathrm{BSF}}
\newcommand{\gBNF}{\gamma_\mathrm{BNF}}
\newcommand{\gBM}{\gamma_\mathrm{BM}}
\newcommand{\gSN}{\gamma_\mathrm{SN}}
\newcommand{\gSF}{\gamma_\mathrm{SF}}

\newcommand{\xiN}{\xi_\mathrm{N}}
\newcommand{\xiF}{\xi_\mathrm{F}}
\newcommand{\PhiN}{\Phi_\mathrm{N}}
\newcommand{\PhiF}{\Phi_\mathrm{F}}
\newcommand{\PhiNF}{\Phi_\mathrm{NF}}
\newcommand{\PhiFdir}{\Phi_\mathrm{F,dir}}
\newcommand{\GFdir}{G_\mathrm{F,dir}}
\newcommand{\PhiS}{\Phi_\mathrm{S}}
\newcommand{\GN}{G_\mathrm{N}}
\newcommand{\GF}{G_\mathrm{F}}
\newcommand{\GNF}{G_\mathrm{NF}}
\newcommand{\GS}{G_\mathrm{S}}
\newcommand{\rhoN}{\rho_\mathrm{N}}
\newcommand{\rhoF}{\rho_\mathrm{F}}

\newcommand{\AN}{A_\mathrm{N}}
\newcommand{\AFdir}{A_\mathrm{F,dir}}
\newcommand{\AF}{A_\mathrm{F}}
\newcommand{\ANF}{A_\mathrm{NF}}
\newcommand{\BN}{B_\mathrm{N}}
\newcommand{\BFdir}{B_\mathrm{F,dir}}
\newcommand{\BNF}{B_\mathrm{NF}}

\newcommand{\PartDeriv}[2]{\frac{\partial #1}{\partial #2}}
\newcommand{\eqref}[1]{(\ref{#1})}
\newcommand{\units}[1]{\ensuremath{\,\mathrm{#1}}}

\usepackage[normalem]{ulem}

\begin{document}

\title[Ferromagnetic planar Josephson junction with transparent interfaces]
{Ferromagnetic planar Josephson junction with transparent interfaces: a $\varphi$ junction proposal}

\author{D~M~Heim$^1$, N~G~Pugach$^{2,3}$, M~Yu~Kupriyanov$^2$, E~Goldobin$^4$, D~Koelle$^4$ and R~Kleiner$^4$}

\address{$^1$ Institut f{\"u}r Quantenphysik and Center for Integrated Quantum Science and Technology (IQ$^\mathrm{ST}$), Universit{\"a}t Ulm, D-89069 Ulm, Germany}

\address{$^2$ Skobeltsyn Institute of Nuclear Physics, M.~V. Lomonosov Moscow State University, 119991 Leninskie Gory, Moscow, Russia}

\address{$^3$ Faculty of Physics, M.~V. Lomonosov Moscow State University, 119991 Leninskie Gory, Moscow, Russia}

\address{$^4$ Physikalisches Institut and Center for Collective Quantum Phenomena in LISA$^+$, Universit\"{a}t T\"{u}bingen, D-72076 T\"{u}bingen, Germany}

\ead{dennis.heim@uni-ulm.de}

\begin{abstract}
We calculate the current phase relation of a planar Josephson junction with a ferromagnetic weak link located on {top of} a thin normal metal film. {Following experimental observations we assume transparent superconductor-ferromagnet interfaces}. This provides the best interlayer coupling and a low suppression of the superconducting correlations penetrating from the superconducting electrodes into the ferromagnetic layer. We show that this Josephson junction is a promising candidate for an experimental $\varphi$ junction realisation.
\end{abstract}


\maketitle

\section{Introduction}\label{sec:introcuction}

A $\varphi$ junction~\cite{mints:1998,buzdin:2003} is a Josephson junction with {a doubly degenerate} ground state, in which the Josephson phase takes the values $+\varphi$ or $-\varphi$ ($0<\varphi<\pi$)~\cite{goldobin:2007}. This junction being closed into a ring is able to self-generate a fractional flux $\Phi_0\varphi$/(2$\pi$), where $\Phi_0$ is the magnetic flux quantum.

{In this sense the $\varphi$ junction is a generalisation of the $\pi$ junction~\cite{bulaevskii:1977} which has a Josephson phase $+\pi$ or $-\pi$ in its ground state. It has been experimentally demonstrated that the $\pi$ junction improves the performance and simplifies the design of classical and quantum circuits~\cite{ustinov:2003,ortlepp:2006,feofanov:2010}. Since the $\varphi$ junction offers the possibility to choose a special value of the phase in the ground state it may further optimize these circuits.}

{The initial $\varphi$ junction proposal \cite{mints:1998} investigated grain-boundary junctions, which were analysed experimentally in \cite{ilichev:1999}. From then on $\varphi$ junctions were studied more and more intensively and many other systems appeared as possible candidates for the realisation of $\varphi$ junctions, e.g. \cite{buzdin:2003,goldobin:2007,cleuziou:2006,gumann:2009,pugach:2010,goldobin:2011,lipman:2012,alidoust:2013}. Only recently,} an experimental evidence of a $\varphi$ junction made of $0$ and $\pi$ parts~\cite{buzdin:2003,pugach:2010,goldobin:2011} was reported~\cite{sickinger:2012}. One half of the junction had the Josephson phase $0$ in its ground state and the other half the phase $\pi$. This was realised~\cite{sickinger:2012} by connecting two superconductor-insulator-ferromagnet-superconductor (SIFS) junctions in parallel. The advantage of this concept is that it is based on the technology already developed for the fabrication of $0$-$\pi$ junctions~\cite{hilgenkamp:2004,
weides:2006}.

\begin{figure}[b]
\begin{center}
\includegraphics{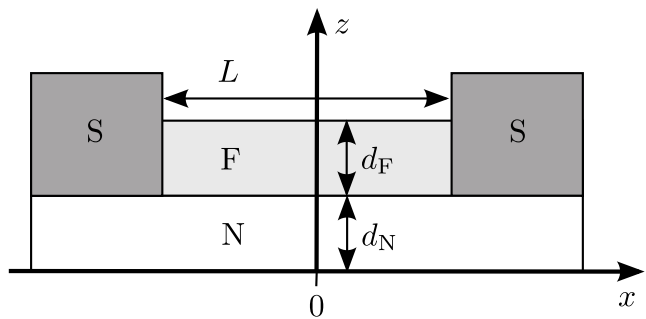}
  \caption{The geometry of the considered system. {The} Josephson junction {consists of two} superconducting (S) electrodes {separated by} a ferromagnetic (F) weak link of thickness $\dFt$ and length $L$. It is located on {top of} a thin normal (N) metal film of thickness $\dNt$.}
  \label{fig:junction_02}
\end{center}
\end{figure}

{On the other hand this $\varphi$ junction concept is difficult to realise experimentally because, e.g., a step in the thickness of the F layer must be realised with very high accuracy~\cite{pugach:2010,goldobin:2011,sickinger:2012}. A completely other method, the ``ramp-type overlap'' (RTO) $\varphi$ junction, was proposed by Bakurskiy et al.~\cite{bakurskiy:2012}. It only requires one small SFS junction located on a thin normal (N) metal layer, see figure~\ref{fig:junction_02}. This basic setup provides a miniaturized $\varphi$ junction. Moreover, this type of junction has {already} been realised experimentally for the analysis of the double proximity effect~\cite{golikova:2012}.}

A simple model~\cite{goldobin:2007} to show that the RTO junction can be used as a $\varphi$ junction requires its current-phase relation (CPR). By writing it in terms of a sine series
\begin{equation}
  I(\phi)=A \sin(\phi)+B \sin(2\phi)
  , \label{eq:cpr:introcuction}
\end{equation}
where $\phi$ is the Josephson phase, the amplitudes have to obey the conditions~\cite{goldobin:2007}
\begin{equation}
  |B|>|A|/2 \quad \mathrm{and} \quad B<0
  . \label{eq:phijunction:condition:a}
\end{equation}

{The RTO junction, schematically shown in figure~\ref{fig:junction_02},} can fulfil these conditions because the current flows between the S electrodes through the F metal \emph{and} the N layer. In this way the properties of an SFS and SNS junction are combined. The SFS junction can have a negative~\cite{golubov:2004,buzdin:2005} amplitude $\AF$ {in~\eqref{eq:cpr:introcuction}}, while the SNS junction has a positive~\cite{golubov:2004,likharev:1979} amplitude $\AN$ {in~\eqref{eq:cpr:introcuction}}. By adding both the total amplitude $A$ can be minimized and a dominant negative amplitude $B$ {from the SNS part} is obtained to fulfil conditions~\eqref{eq:phijunction:condition:a}. Since supercurrents in SFS junctions are rather small, the SNS contribution has to be reduced. This is done by using only a thin normal metal film.

{In the present paper we investigate an RTO junction which has, differently from the one proposed in~\cite{bakurskiy:2012}, transparent SF interfaces in order to amplify the SFS contribution to the total current. This assumption {has} already successfully {been} used to describe various experiments~\cite{golikova:2012,oboznov:2006,bannykh:2009}. As a result, we obtain slightly {smaller} system sizes for the $\varphi$ junction realisation than~\cite{bakurskiy:2012}, where weakly transparent interfaces were assumed. Moreover, our approach provides a better penetration of the superconducting correlations into the F layer which may increase the Josephson current. In the framework of transparent SF interfaces we cannot use linearised equations for the SFS part, as it was done in~\cite{bakurskiy:2012}. Therefore, we use non-linearised equations in the SFS \emph{and} SNS part for our analytical approach.}

We derive the CPR in the ``dirty'' limit. For this purpose, we combine the solution of the Usadel equations in the N film~\cite{bakurskiy:2012} with the solution of the Usadel equations in the SFS layer~\cite{buzdin:1991}. The resulting current phase relation  consists of three parts: (i) a contribution from the SFS layer, (ii) a contribution from the N film and (iii) a composite SNFS term.

The paper is organized as follows. In section~\ref{sec:model} we introduce the model of the considered Josephson junction in terms of Usadel equations. The analytical expression of the CPR of our system is based on this model and presented in section~\ref{sec:calculation}. In section~\ref{sec:discussion} we use this expression together with realistic system parameters to discuss its applicability as $\varphi$ junction. Finally, an appendix provides a detailed derivation of the composite SNFS current.

\section{Model}\label{sec:model}

The considered Josephson junction is sketched in figure~\ref{fig:junction_02}. It consists of an SFS junction located on a normal metal film. The F layer has a thickness $\dFt$ and a length $L$ while the N layer has a thickness $\dNt$ and is considered as infinitely long. We have chosen the $x$ and $z$ axis in directions {parallel and perpendicular} to the plane of the N film, respectively.

For the calculation of the current $I(\phi)$ flowing from one superconducting electrode to the other we determine the Green's functions describing our system. {We consider} the ``dirty'' limit \cite{golikova:2012,oboznov:2006,bannykh:2009}, in which the elastic scattering length is much smaller than the characteristic decay length, we can use the Usadel equations~\cite{usadel:1970} to model our system. We write them in the form~\cite{golubov:2004}
\begin{eqnarray}
  \frac{\xi_j^2}{G_j}\left[\frac{\partial}{\partial x} \left( G_j^2 \frac{\partial}{\partial x} \Phi_j \right)
  +\frac{\partial}{\partial z} \left( G_j^2 \frac{\partial}{\partial z} \Phi_j \right) \right]
  - \frac{\omegat}{\pi \Tc} \Phi_j = 0, \nonumber \\
  G_j=\frac{\omegat}{\sqrt{\omegat^2+\Phi_j\Phi_j^*}}, \quad j \in \{\mathrm{N,F}\}
  \label{eq:usadel:general}
\end{eqnarray}
in the N and F layer, respectively. Here, $\Phi_j$ and $G_j$ are the Usadel Green's functions in the $\Phi$ parametrization~\cite{kupriyanov:1988}. The frequencies $\omegat=\omega+\ii H$ contain the Matsubara frequencies $\omega=\pi T (2n+1)$ {at temperature $T$}, where $n=0,1,2,\ldots$, and the exchange field $H$ of the ferromagnetic material which is assumed to be zero in the N layer. The decay lengths
\begin{equation}
  \xiN=\sqrt{\frac{D_\mathrm{N}}{2\pi \Tc}}, \quad \xiF=\sqrt{\frac{D_\mathrm{F}}{2\pi \Tc}}
\end{equation}
of the superconducting correlations are defined via the critical temperature $\Tc$ of the superconductor (we {use} $\hbar=k_\mathrm{B}=1$) and the diffusion coefficients $D_\mathrm{N}$ and $D_\mathrm{F}$ in the normal and ferromagnetic metal, respectively.

We assume that superconductivity {in the S electrodes} is not suppressed by the neighbouring N and F {layers}. This assumption is valid in our case of transparent SF interfaces {with} the conditions for the suppression parameters
\begin{eqnarray}
 \gBSF=\frac{R_\mathrm{BSF}A_\mathrm{BSF}}{\rho_\mathrm{F}\xi_\mathrm{F}}\ll 1,\quad \gSF=\frac{\rho_\mathrm{S}\xi_\mathrm{S}}{\rho_\mathrm{F}\xi_\mathrm{F}} \ll 1, \\
 \gBSN=\frac{R_\mathrm{BSN}A_\mathrm{BSN}}{\rho_\mathrm{N}\xi_\mathrm{N}}\gg \gSN=\frac{\rho_\mathrm{S}\xi_\mathrm{S}}{\rho_\mathrm{N}\xi_\mathrm{N}}
 . \label{eq:gamma:bsn}
\end{eqnarray}
Here, $R_\mathrm{BSN,BSF}$ and $A_\mathrm{BSN,BSF}$ are the resistances and areas of the SN and SF interfaces. The values of $\rho_\mathrm{N,F,S}$ describe the resistivity of the N, F, and S metals.

This allows us to use the rigid boundary conditions~\cite{golubov:2004}
\begin{equation}
 \Phi_\mathrm{S}(\pm L/2)=\Delta \exp(\pm\ii \phi/2), \quad G_\mathrm{S}=\frac{\omega}{\sqrt{\omega^2+\Delta^2}}
 , \label{eq:bc:rigid}
\end{equation}
where $\Delta$ is the absolute value of the order parameter in the superconductor.

The boundary conditions~\cite{kupriyanov:1988,koshina:2000,golubov:2004} at the free interfaces are
\begin{equation}
 \frac{\partial}{\partial z} \Phi_j=0,\quad j \in \{\mathrm{N,F}\}
  , \label{eq:bc:free}
\end{equation}
and at the interfaces of the superconductor they are
\begin{eqnarray}
 \gBSN \xiN \PartDeriv{\PhiN}{z}=\frac{\GS}{\GN} \left[\PhiS(\pm L/2) - \PhiN \right]
 \label{eq:bc:SN}
\end{eqnarray}
and
\begin{eqnarray}
 \PhiF=\frac{\omegat}{\omega}\PhiS(\pm L/2)
 . \label{eq:bc:SF}
\end{eqnarray}
Additionally we use
\begin{equation}
 \gBNF \xiF \PartDeriv{\PhiF}{z} = \frac{\GN}{\GF}\left( \frac{\omegat}{\omega} \PhiN - \PhiF \right)
 \label{eq:bc:NF}
\end{equation}
at the NF interfaces, where
\begin{equation}
 \gBNF=\frac{R_\mathrm{BNF}A_\mathrm{BNF}}{\rho_\mathrm{F}\xi_\mathrm{F}}
 \label{eq:gamma:bnf}
\end{equation}
is defined analogous to~\eqref{eq:gamma:bsn}.

Finally we calculate the total current
\begin{equation}
  I(\phi)=I_\mathrm{N}(\phi)+I_\mathrm{F}(\phi)
  \label{eq:current:tot:def}
\end{equation}
by integrating the standard expressions~\cite{golubov:2004} for the current densities of the N and F part over the junction cross section along the $z$ axis. This leads us to
\begin{eqnarray}
 I_\mathrm{N}(\phi)&=&\ii \frac{\pi T W}{2 e \rhoN} \sum_{\omega=-\infty}^{\infty} \int_{0}^{\dNt} \dd z \frac{\GN^2}{\omega^2} \nonumber \\
 && \times \left[ \PhiN(\omega)\ \PartDeriv{}{x} \PhiN^*(-\omega) - \PhiN^*(-\omega)\ \PartDeriv{}{x} \PhiN(\omega) \right]_{x=0}
 \label{eq:current:n:tot:def}
\end{eqnarray}
and
\begin{eqnarray}
 I_\mathrm{F}(\phi)&=&\ii \frac{\pi T W}{2 e \rhoF} \sum_{\omega=-\infty}^{\infty} \int_{\dNt}^{\dNt+\dFt} \dd z \frac{\GF^2}{\omegat^2} \nonumber \\
 && \times \left[ \PhiF(\omega)\ \PartDeriv{}{x} \PhiF^*(-\omega) - \PhiF^*(-\omega)\ \PartDeriv{}{x} \PhiF(\omega) \right]_{x=0}
 . \label{eq:current:f:tot:def}
\end{eqnarray}
The width $W$ of the junction {along the $y$ axis} is supposed to be small compared to the Josephson penetration depth. We have chosen the position $x=0$ for the integration over the junction cross section since the $z$ component of the current densities vanishes there because of the symmetry of the considered {junction geometry}.

\section{Currents}\label{sec:calculation}

In order to calculate the current $I(\phi)$ from~\eqref{eq:current:tot:def} we cannot simply add the current through the N layer calculated by Bakurskiy et al.~\cite{bakurskiy:2012} to the SFS current calculated by Buzdin et al.~\cite{buzdin:1991} because we have to take into account a composite SNFS current which appears due to a penetration of superconductivity from the N layer into the F layer. Therefore, we split the current $I_\mathrm{F}(\phi)$ into a contribution $I_\mathrm{F,dir}(\phi)$ due to a direct penetration of superconductivity into the F layer and the additional part $I_\mathrm{NF}(\phi)$. This leads us to
\begin{equation}
 I(\phi)=I_\mathrm{N}(\phi)+I_\mathrm{F,dir}(\phi)+I_\mathrm{NF}(\phi)
 . \label{eq:current:tot:def:new}
\end{equation}
In the following three sections we derive the expressions of these three currents using the scaling
\begin{equation}
 \widetilde{I}_{j}(\phi) = I_{}(\phi) \frac{e \rho_j}{W \Delta }
 . \label{eq:current:scaling}
\end{equation}

\subsection{Current in the N layer} \label{sec:n-part}

In this layer we adopt the current
\begin{eqnarray}
  \widetilde{I}_\mathrm{N}(\phi)
    = 2 \frac{\dNt T}{\xiN \Tc} \sum_{\omega>0} \frac{\Gamma(\phi)}{\mu(\phi)}\, r \sin(\phi)
       \label{eq:cpr:sum:n}
\end{eqnarray}
with the definitions
  \begin{eqnarray}
    \Gamma(\phi) = \frac{r \delta \sqrt{\gBM \Omega + \GS}}
    { \sqrt{2 \gBM \Omega (\sqrt{\Omega^2+\delta^2 r^2} + \mu(\phi))}}, \label{eq:def:gamma_phi} \\
    \delta = \frac{\Delta}{\pi \Tc}, \quad \gBM = \frac{\gBSN \dNt}{\xiN}, \quad \Omega=\frac{\omega}{\pi \Tc}, \\
    r = \left( \frac{\gBM}{\pi \Tc} \sqrt{\omega^2 + \Delta^2 }+1 \right)^{-1}, \label{eq:def:r}\\
    \mu(\phi) = \sqrt{ \Omega^2 + r^2 \delta^2 \cos^2 ({\phi}/{2}) } \label{eq:def:mu}
     \label{eq:functions:n:sinphi}
 \end{eqnarray}
 \label{eq:functions:n:def}
from~\cite{bakurskiy:2012}. Its derivation is based on the assumption $L \ll \xiN$ and an infinitely long N layer. It is calculated with the help of the solution $\PhiN(x)$~\eqref{eq:green:n} of the non-linear Usadel equations which depends only on the coordinate $x$ because the thickness $\dNt \ll \xiN$ is assumed to be small.

\subsection{Current in the F layer} \label{sec:f-part}

The current
  \begin{eqnarray}
  I_\mathrm{F,dir}(\phi) = \sqrt{2}\,64 {\dFt} \kappa\, \ee^{-2\kappa L} \mathcal{F} \sin\left( 2\kappa L + \frac{\pi}{4} \right)\sin\phi
      , \label{eq:cpr:f}
  \end{eqnarray}
with
\begin{eqnarray}
 \kappa=\frac{\sqrt{h}}{\sqrt{2} \xiF}, \quad { h=\frac{H}{\pi \Tc}, }\quad \mathcal{F}= \pi T \sum_{\omega>0} \frac{\Theta^2}{\Delta}, \label{eq:def:kappa} \\
 \Theta = \frac{\Delta}{\eta+|\omega|+\sqrt{2\eta(\eta+|\omega|)}}, \quad \eta=\sqrt{\omega^2+\Delta^2}, \label{eq:def:eta}
\end{eqnarray}
is a result of~\cite{buzdin:1991}. It also has been calculated with the help of a solution of the non-linear Usadel equations because $\gBSF=0$ is assumed. Additionally the condition $\xiF \ll L$ is required.

\subsection{Composite NF current}\label{sec:fn-part}

We determine the current $I_\mathrm{NF}(\phi)$ by combining the two non-linear solutions $\PhiFdir(x)$ and $\PhiN(x)$ of~\eqref{eq:green:F} and~\eqref{eq:green:n} in~\ref{sec:nf:current:deriv}. The main idea is to decompose the ferromagnetic Green's function
\begin{equation}
 \PhiF(x,z) = \PhiFdir(x) + \PhiNF(x,z)
  \label{eq:green:F:total}
\end{equation}
into a function $\PhiFdir(x)$, which corresponds to currents only flowing in the F layer, and a function $\PhiNF(x,z)$, which corresponds to currents flowing through the N layer into the F layer. The second function is obtained by linearising the Usadel equations~\eqref{eq:usadel:general} in the F {layer}. Then we connect it to the N layer solution $\PhiN(x)$ via the boundary conditions.

The superposition~\eqref{eq:green:F:total} of the solution $\PhiFdir$ of the non-linear Usadel equation with the solution $\PhiNF$ of the linearised Usadel equation is valid because we distinguish in the F part between two cases: (i) at $x \approx \pm L / 2$ near the boundaries to the S regions the Green's function $\PhiFdir$ is very dominant $\left|\PhiFdir \right| \gg \left| \PhiNF \right|$  due to a transparent boundary between the S and the F part, that is $\gBSF=0$; (ii) at $x \approx 0$, that is away from the boundaries the contribution of $\PhiF$ decays exponentially. Therefore, the contribution from the N part is dominant $\left| \PhiNF \right| \gg \left| \PhiFdir \right|$.

As a result~\eqref{eq:cpr:sum:fn} we obtain the current
  \begin{eqnarray}
  \widetilde{I}_{NF}(\phi)
  = \frac{16 \cos(\phi/2) \xiF}{\gBNF h \Delta \xiN} \ee^{-\kappa L/2} \nonumber \\
  \times \left[ \sin\frac{\kappa L}{2} + \frac{\kappa  L}{\sqrt{2}} \ee^{-\kappa L/2} \cos \left(\kappa L+\frac{\pi}{4}\right) \right] \nonumber \\
   \times 2\pi T \sum_{\omega>0} \Theta\, \Gamma(\phi) \sin \frac{\phi}{2}
  ,  \label{eq:cpr:sum:fn:main}
  \end{eqnarray}
with the definitions of $\Gamma(\phi)$ from~\eqref{eq:def:gamma_phi}, $\kappa$ from~\eqref{eq:def:kappa} and $\Theta$ together with $\eta$ from~\eqref{eq:def:eta}.

\section{Discussion}\label{sec:discussion}

In this section we estimate the geometrical parameters $\dNt$, $\dFt$ and $L$, see figure~\ref{fig:junction_02}, for which the considered Josephson junction obeys the $\varphi$ junction conditions~\eqref{eq:phijunction:condition:a}. We use the analysing scheme of~\cite{bakurskiy:2012} and finally compare our results with the ones {obtained in~\cite{bakurskiy:2012}}.

We split the sine series amplitudes
\begin{equation}
  A=\AN+\AFdir+\ANF
  , \label{eq:A}
\end{equation}
\begin{equation}
  B=\BN+\BNF
  \label{eq:B}
\end{equation}
of the total current~\eqref{eq:current:tot:def:new}, scaled according to~\eqref{eq:current:scaling}, into parts originating from the current of the N layer~\eqref{eq:cpr:sum:n}, the F layer~\eqref{eq:cpr:f} and the composite NF current~\eqref{eq:cpr:sum:fn:main}. There is no amplitude $\BFdir$ because we have {a} pure sinusoidal CPR~\eqref{eq:cpr:f} in the F layer.

In our calculations we chose the temperature $T=0.1\,\Tc$. We make this choice because far away from the critical temperature the CPR has larger deviations from the $\sin\phi$ form~\cite{likharev:1979} which results in a larger second harmonic $B$. As S electrode material we chose Nb with $\Tc=9.2\units{K}$ because it is commonly used in superconducting circuits.

\begin{figure}[ht]
\begin{center}
\includegraphics{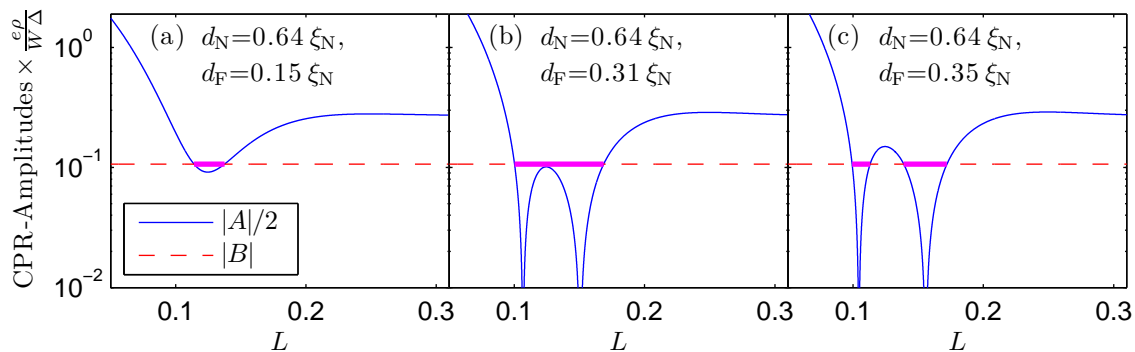}
  \caption{The functions $|A|/2$ and $|B|$, based on~\eqref{eq:A} and~\eqref{eq:B}, as functions of $L$ for $\dNt=0.\dN\,\xiN$ and three characteristic values of $\dFt$. The bold lines correspond to values of $L$ where the conditions~\eqref{eq:phijunction:condition:a} for the $\varphi$ junction realization are fulfilled.}
  \label{fig:cpr:total}
\end{center}
\end{figure}

Our first step is to find suitable parameters $\dFt$. For this purpose we analyse the amplitudes~\eqref{eq:A} and~\eqref{eq:B} as a function of $L$ for different values of $\dFt$ for {the same parameters as in~\cite{bakurskiy:2012}:} $\dNt=0.\dN\,\xiN$, $\xi_F=0.1\,\xi_N$, $H=10\,\Tc$, $\Delta=1.76\,\Tc$, $\rhoF=\rhoN=\rho$ and $\gBNF=1$. Figure~\ref{fig:cpr:total} shows three typical examples: (a) $\dFt=0.\dFa\,\xi_N$, (b) $\dFt=0.31\,\xi_N$ and (c) $\dFt=0.\dFb\,\xi_N$. The first (a) and last (c) examples correspond to limiting cases where it is difficult to realize a $\varphi$ junction because the intervals of $L$ where conditions~\eqref{eq:phijunction:condition:a} hold are not large. These intervals of $L$ are highlighted by bold lines. In between the two limiting values for $\dFt$ this line becomes longer. Figure~\ref{fig:cpr:total} (b) shows an optimum situation because there is a wide range {of} $L$ {which yields} a $\varphi$ junction
configuration.

For the optimum value $\dFt=0.31\,\xi_N$ we calculate the magnitudes \mbox{$\Upsilon_A=\AN{W \Delta }/{(e \rho)}  = 0.534$} and \mbox{$\Upsilon_B= {\BN} {W \Delta }/{(e \rho)}  = -0.106$}. Inserting them together with the amplitude $\AFdir$ from~\eqref{eq:cpr:f} into~\eqref{eq:phijunction:condition:a} and neglecting the small NF contributions leads us to the condition
\begin{equation}
   \left| \Upsilon_A +\frac{1}{\varepsilon} \Psi(L) \right| < 2 \left| \Upsilon_B \right|
   . \label{eq:phijunction:condition:disc}
\end{equation}
Here, we use the constant $\varepsilon={\xiF}/{(64 \mathcal{F} \sqrt{h} \dFt)}$ with $\dFt=0.31\,\xi_N$, $\mathcal{F}=0.0691$ and
\begin{equation}
 \Psi(L)=\exp(-2\kappa L)\sin\left(2\kappa L+ \pi/4 \right)
 .
\end{equation}
From~\eqref{eq:phijunction:condition:disc} we find the minimum value $0.\La\,\xiN$ and maximum values $0.\Lb\,\xiN$ of $L$.

For summarising our suggestion of the geometrical configuration of a $\varphi$ junction we use the value $\xiN=100\units{nm}$ for Cu as N layer, a strongly diluted ferromagnet such as FePd or the CuNi alloy with $\xiN=10\units{nm}$ and $H = 10\,\Tc$ as F metal. Our set of parameters then become \mbox{$\dNt \gtrsim 50 \units{nm}$}, \mbox{$\dFa \units{nm} \lesssim \dFt \lesssim \dFb \units{nm}$} and \mbox{$\La \units{nm} \lesssim L \lesssim \Lb \units{nm}$}, which we compare to the values \mbox{$\dNt \gtrsim 50 \units{nm}$}, \mbox{$19 \units{nm} \lesssim \dFt \lesssim 48 \units{nm}$} and \mbox{$7 \units{nm} \lesssim L \lesssim 22 \units{nm}$} of~\cite{bakurskiy_note}.

Since we use the same N layer configuration, the value for $\dNt$ is the same. But the suggested regime for $\dFt$ differs. A change in this direction was expected because we only need a thin F layer since the transparency of our interfaces already amplifies our SFS current contribution. The possible range for the length $L$ of the F part is smaller in our case but the whole junction configuration is still experimentally feasible.

\section{Conclusion}\label{sec:conclusion}

We have shown that the considered Josephson junction with a ferromagnetic weak link located on a thin normal metal film is a good candidate for a $\varphi$ junction realisation. By choosing transparent SF interfaces we obtained slightly different system sizes {for the $\varphi$ junction existence} compared to a junction with weakly transparent interfaces.

The current was split into a contribution through the N layer, the F layer and a composite term which described the current flowing through the N and F parts of the junction simultaneously. We performed our calculations in the ``dirty'' limit, that is, the currents are {obtained from} solutions of the non-linear Usadel equations.

Since our case of a large interface transparency corresponds better to the experimental situation~\cite{oboznov:2006,bannykh:2009,golikova:2012} {than weakly transparent interfaces~\cite{bakurskiy_note}} it is important to note that a smaller thickness and length of {the} F layer have to be chosen than predicted in~\cite{bakurskiy_note}. We are looking forward to experiments realising this $\varphi$ junction and its application in classical and quantum devices.

\ack
  We thank S V Bakurskiy for fruitful and stimulating discussions. DMH thanks Professor W P Schleich and K Vogel for giving him the possibility to work at the M.~V. Lomonosov Moscow State University. Financial support by the DFG (Project No. SFB/TRR-21), the Russian Foundation for Basic Researches (RFBR grants No. 11-02-12065-ofi-m, 13-02-01452-a) and the Ministry of Education and Science of the Russian Federation (grant 8641) is gratefully acknowledged.

\appendix*

\section{NF current derivation} \label{sec:nf:current:deriv}

In this appendix we derive the current~\eqref{eq:cpr:sum:fn:main} which flows through the N and F part of the junction, sketched in figure~\ref{fig:junction_02}, simultaneously. We first linearise the Usadel equations~\eqref{eq:usadel:general} and then combine the solution with the Green's functions from~\cite{bakurskiy:2012} and~\cite{buzdin:1991}.

For the linearisation of~\eqref{eq:usadel:general} we assume the superconducting correlation coming from the N part into the F part as rather small. Then, the Green's function $\PhiNF(x,z)$ can also be assumed to be small. Using $\GNF=\sign(\omega)$ we obtain the linearised Usadel equation~\cite{buzdin:2005}
\begin{equation}
 \xiF^2 \left(\frac{\partial^2}{\partial x^2} + \frac{\partial^2}{\partial z^2} \right) \PhiNF = \Omegat\ \PhiNF
 , \label{eq:usade:linearized}
\end{equation}
with the definitions
\begin{equation}
 \Omegat = |\Omega|+\ii\ \sign(\Omega) h, \quad \Omega=\frac{\omega}{\pi \Tc}, \quad  h=\frac{H}{\pi \Tc}
 .
\end{equation}

Its solution in the form of a series
\begin{equation}
  \PhiNF(x,z)= \sum_{n=1}^{\infty} b_n \sin \left( \frac{2\pi}{L} n x\right) \cosh \left[ \kappa_n (z-\dNt-\dFt) \right]
  , \label{eq:green:NF}
\end{equation}
with
\begin{equation}
 \kappa_n^2 = \left( \frac{2\pi n}{L} \right)^2 + \frac{\Omegat}{\xiF^2}
\end{equation}
and a Fourier coefficient $b_n$, already obeys the boundary condition~\eqref{eq:bc:free} at the upper border ($z=\dNt+\dFt$).

The boundary conditions at the left and right end of the F part at $x=\pm L/2$ are also already fulfilled. They follow from~\eqref{eq:bc:SF} with $\gBSF=0$. Using here the definition~\eqref{eq:green:F:total} of $\PhiF$ leads us to the condition
\begin{equation}
 \PhiFdir(\pm L/2) + \PhiNF= \frac{\Omegat}{\Omega}\PhiS(\pm L/2)
 .
\end{equation}
This equation is already fulfilled by the solution
 \begin{equation}
  \PhiFdir(x)=\frac{\Omegat}{\GFdir} \left( \ee^{-\ii \phi/2}\sin\alpha^- + \ee^{+\ii \phi/2}\sin\alpha^+ \right)
   \label{eq:green:F}
 \end{equation}
 with
\begin{eqnarray}
 \alpha^{\pm} &=& 4 \arctan\left[ \Theta \exp \left( \pm \sqrt{\widetilde\Omega}\ \frac{x\mp L/2}{\xiF} \right) \right], \\
 \Theta &=& \frac{\Delta}{\eta+|\omega|+\sqrt{2\eta(\eta+|\omega|)}}, \label{eq:functions:f:a}
\end{eqnarray}
from~\cite{buzdin:1991} alone. Therefore, the NF Green's function~\eqref{eq:green:NF} only has to obey the conditions $\PhiNF=0$ at $x=\pm L/2$. Note that we do not need the expression for $\GFdir$ to finally calculate the current.

In order to obtain the Fourier coefficient $b_n$ which fixes the solution $\PhiNF$ from~\eqref{eq:green:NF} we use the boundary condition~\eqref{eq:bc:NF} where we neglect the term $\PhiNF$ assuming $|\PhiNF|\ll|\PhiN|$. Now, we replace the Green's function $\GF$ by $\GNF=\sign(\omega)$ and insert
 \begin{eqnarray}
    \PhiN(x) = r \Delta \cos \frac{\phi}{2} + 2 \ii \mu(\phi) \Gamma(\phi) \sin \left(\frac{\phi}{2}\right) \frac{x}{\xiN},\quad
    \GN = \frac{\Omega}{\mu(\phi)},
    \label{eq:green:n}
 \end{eqnarray}
 from~\cite{bakurskiy:2012}, where we use the definitions~\eqref{eq:def:gamma_phi},~\eqref{eq:def:r} and~\eqref{eq:def:mu} for $\Gamma(\phi)$, $r$ and $\mu(\phi)$, respectively.  By neglecting the real part of $\PhiN$ we obtain the Fourier coefficient
\begin{equation}
 b_n = \frac{2\ii\, \Omegat L (-1)^n \Gamma(\phi) \sin(\phi/2)}{\kappa_n \sinh(\kappa_n \dFt)\, \xiF\, \gBNF \pi n \,\xiN}
 .
\end{equation}

Our last step is to calculate the current $I_\mathrm{NF}$. Therefore, we insert the Green's function~\eqref{eq:green:F:total}, which contains the Green's functions from~\eqref{eq:green:NF} and~\eqref{eq:green:F}, into the definition~\eqref{eq:current:f:tot:def} of the F layer current. Due to the condition $x=0$ it reduces to a sum $I_\mathrm{F}(\phi)=I_\mathrm{NF}(\phi)+I_\mathrm{F,dir}(\phi)$, where the NF current is defined by
\begin{eqnarray}
 I_\mathrm{NF}(\phi)&=&\ii \frac{\pi T W}{2 e \rhoF} \sum_{\omega=-\infty}^{\infty} \int_{\dNt}^{\dNt+\dFt} \dd z \frac{\GNF \GFdir}{\omegat^2} \nonumber \\
 && \times \left[ \PhiFdir(\omega)\ \PartDeriv{}{x} \PhiNF^*(-\omega) - \PhiFdir^*(-\omega)\ \PartDeriv{}{x} \PhiNF(\omega) \right]_{x=0}
 \label{eq:I_NF:def}
\end{eqnarray}
and $I_\mathrm{F,dir}(\phi)$ is the current flowing only through the F layer~\cite{buzdin:1991} summarized in~\eqref{eq:cpr:f}.

We insert the Green's functions $\PhiNF$ and $\PhiFdir$ from~\eqref{eq:green:NF} and~\eqref{eq:green:F} into~\eqref{eq:I_NF:def}. Then, by finally using the the approximation $\Omegat \approx \ii h$, which holds for the condition $\pi \Tc \ll H$, we obtain the scaled current
  \begin{eqnarray}
  \widetilde{I}_{NF}(\phi)
  = \frac{16 \cos(\phi/2) \xiF}{\gBNF h \Delta \xiN} \ee^{-\kappa L/2} \nonumber \\
  \times \left[ \sin\frac{\kappa L}{2} + \frac{\kappa  L}{\sqrt{2}} \ee^{-\kappa L/2} \cos \left(\kappa L+\frac{\pi}{4}\right) \right] \nonumber \\
   \times 2\pi T \sum_{\omega>0} \Theta\, \Gamma(\phi) \sin \frac{\phi}{2}
  .  \label{eq:cpr:sum:fn}
  \end{eqnarray}

\section*{References}

\providecommand{\newblock}{}


\begin{thebibliography}{10}
\expandafter\ifx\csname url\endcsname\relax
  \def\url#1{{\tt #1}}\fi
\expandafter\ifx\csname urlprefix\endcsname\relax\def\urlprefix{URL }\fi
\providecommand{\eprint}[2][]{\url{#2}}

\bibitem{mints:1998}
Mints R~G 1998 {\em Phys. Rev. B\/} {\bf 57} R3221--R3224

\bibitem{buzdin:2003}
Buzdin A and Koshelev A~E 2003 {\em Phys. Rev. B\/} {\bf 67} 220504

\bibitem{goldobin:2007}
Goldobin E, Koelle D, Kleiner R and Buzdin A 2007 {\em Phys. Rev. B\/} {\bf 76}
  224523

\bibitem{bulaevskii:1977}
Bulaevskii L~N, Kuzii V~V and Sobyanin A~A 1977 {\em Pis'ma Zh. Eksp. Teor.
  Phys.\/} {\bf 25} 314--318 [1977 \textit{JETP Lett.} \textbf{25} 290--294]

\bibitem{ustinov:2003}
Ustinov A~V and Kaplunenko V~K 2003 {\em J. Appl. Phys.\/} {\bf 94} 5405--5407

\bibitem{ortlepp:2006}
Ortlepp T, Ariando, Mielke O, Verwijs C~J~M, Foo K~F~K, Rogalla H, Uhlmann F~H
  and Hilgenkamp H 2006 {\em Science\/} {\bf 312} 1495--1497

\bibitem{feofanov:2010}
Feofanov A~K, Oboznov V~A, Bol'ginov V~V, Lisenfeld J, Poletto S, Ryazanov V~V,
  Rossolenko A~N, Khabipov M, Balashov D, Zorin A~B, Dmitriev P~N, Koshelets
  V~P and Ustinov A~V 2010 {\em Nature Phys.\/} {\bf 6} 593--597

\bibitem{ilichev:1999}
Il'ichev E, Zakosarenko V, IJsselsteijn R~P~J, Hoenig H~E, Meyer H~G, Fistul
  M~V and M\"uller P 1999 {\em Phys. Rev. B\/} {\bf 59} 11502--11505

\bibitem{cleuziou:2006}
Cleuziou J~P, Wernsdorfer W, Bouchiat V, Ondar\c{c}uhu T and Monthioux M 2006
  {\em Nat. Nano.\/} {\bf 1} 53--59

\bibitem{gumann:2009}
Gumann A and Schopohl N 2009 {\em Phys. Rev. B\/} {\bf 79} 144505

\bibitem{pugach:2010}
Pugach N~G, Goldobin E, Kleiner R and Koelle D 2010 {\em Phys. Rev. B\/} {\bf
  81} 104513

\bibitem{goldobin:2011}
Goldobin E, Koelle D, Kleiner R and Mints R~G 2011 {\em Phys. Rev. Lett.\/}
  {\bf 107} 227001

\bibitem{lipman:2012}
Lipman A, Mints R~G, Kleiner R, Koelle D and Goldobin E 2012
  (\textit{Preprint} \eprint{http://arxiv.org/abs/1208.4057})

\bibitem{alidoust:2013}
Alidoust M and Linder J 2013 {\em Phys. Rev. B\/} {\bf 87} 060503

\bibitem{sickinger:2012}
Sickinger H, Lipman A, Weides M, Mints R~G, Kohlstedt H, Koelle D, Kleiner R
  and Goldobin E 2012 {\em Phys. Rev. Lett.\/} {\bf 109} 107002

\bibitem{hilgenkamp:2004}
Smilde H~J~H, Ariando, Blank D~H~A, Gerritsma G~J, Hilgenkamp H and Rogalla H
  2002 {\em Phys. Rev. Lett.\/} {\bf 88} 057004

\bibitem{weides:2006}
Weides M, Kemmler M, Kohlstedt H, Waser R, Koelle D, Kleiner R and Goldobin E
  2006 {\em Phys. Rev. Lett.\/} {\bf 97} 247001

\bibitem{bakurskiy:2012}
Bakurskiy S~V, Klenov N~V, Karminskaya T~Y, Kupriyanov M~Y and Golubov A~A 2012
  {\em Supercond. Sci. Technol.\/} {\bf 26} 015005

\bibitem{golikova:2012}
Golikova T~E, H\"ubler F, Beckmann D, Batov I~E, Karminskaya T~Y, Kupriyanov
  M~Y, Golubov A~A and Ryazanov V~V 2012 {\em Phys. Rev. B\/} {\bf 86} 064416

\bibitem{golubov:2004}
Golubov A~A, Kupriyanov M~Y and Il'ichev E 2004 {\em Rev. Mod. Phys.\/} {\bf
  76} 411--469

\bibitem{buzdin:2005}
Buzdin A~I 2005 {\em Rev. Mod. Phys.\/} {\bf 77} 935--977

\bibitem{likharev:1979}
Likharev K~K 1979 {\em Rev. Mod. Phys.\/} {\bf 51} 101--159

\bibitem{oboznov:2006}
Oboznov V~A, Bol'ginov V~V, Feofanov A~K, Ryazanov V~V and Buzdin A~I 2006 {\em
  Phys. Rev. Lett.\/} {\bf 96} 197003

\bibitem{bannykh:2009}
Bannykh A~A, Pfeiffer J, Stolyarov V~S, Batov I~E, Ryazanov V~V and Weides M
  2009 {\em Phys. Rev. B\/} {\bf 79} 054501

\bibitem{buzdin:1991}
Buzdin A~I and Kupriyanov M~Y 1991 {\em Pis'ma Zh. Eksp. Teor. Phys.\/} {\bf
  53} 308--312 [1991 \textit{JETP Lett.} \textbf{53} 321--326]

\bibitem{usadel:1970}
Usadel K~D 1970 {\em Phys. Rev. Lett.\/} {\bf 25} 507--509

\bibitem{kupriyanov:1988}
Kuprianov M~Y and Lukichev V~F 1988 {\em Zh. Eksp. Teor. Fiz.\/} {\bf 94}
  139--149 [1988 \textit{Sov. Phys. JETP} \textbf{67} 1163--1168]

\bibitem{koshina:2000}
Koshina E~A and Krivoruchko V~N 2000 {\em Low Temp. Phys.\/} {\bf 26} 115--120

\bibitem{bakurskiy_note}
  The estimations for the thickness $\dFt$ of an RTO $\varphi$ junction with
  weakly transparent SF interfaces are taken from \cite{bakurskiy:2012} and
  divided by a missing factor $\pi$.

\end{thebibliography}
\end{document}